\documentclass[sigconf]{acmart}
\AtBeginDocument{%
  }

\setcopyright{acmlicensed}
\copyrightyear{2025}
\acmYear{2025}
\setcopyright{acmlicensed}
\acmConference[Anonymous]{Anonymous Conference}{Month Day--Day, 2025}{Location}
\acmBooktitle{Proceedings of the Anonymous Conference, Month Day--Day, 2025, Location}

\usepackage{tabularx}
\usepackage{caption}
\usepackage{subcaption}
\usepackage{booktabs}
\usepackage{soul}
\usepackage{multirow}
\usepackage{makecell}
\usepackage{balance}
\usepackage{float}
\usepackage{longtable}
\usepackage{lipsum}
\usepackage{setspace}
\usepackage{subcaption}

\begin{document}

\title[Prevalence, Spreaders, and Responses to AI-Generated Images on $\mathbb{X}$]{From Pixels to Polls: Prevalence, Spreaders, and Responses to AI-Generated Images on $\mathbb{X}$ during the 2024 U.S. Election}
\title[Prevalence, Spreaders, and Responses to AI-Generated Images on $\mathbb{X}$]{From Pixels to Polls: Prevalence, Spreaders, and Responses to AI-Generated Political Images on $\mathbb{X}$}
\title[Prevalence, Spreaders, and Emotional Reception of AI-Generated Political Images on $\mathbb{X}$]{Synthetic Politics: Prevalence, Spreaders, and Emotional Reception of AI-Generated Political Images on $\mathbb{X}$}
%

\author{Zhiyi Chen}
\authornote{These two authors contributed equally to this work.}
\orcid{}
\affiliation{%
  \institution{University of Southern California}
  \city{Los Angeles}
  \state{CA}
  \postcode{90089}
  \country{USA}
}
\email{zchen346@usc.edu}

\author{Jinyi Ye}
\authornotemark[1]
\orcid{0009-0004-7757-1642}
\affiliation{%
  \institution{University of Southern California}
  \city{Los Angeles}
  \state{CA}
  \postcode{90089}
  \country{USA}
}
\email{jinyiy@usc.edu}

\author{Beverlyn Tsai}
\orcid{}
\affiliation{%
  \institution{University of Southern California}
  \city{Los Angeles}
  \state{CA}
  \postcode{90089}
  \country{USA}
}
\email{beverlyn@usc.edu}

\author{Emilio Ferrara}
\orcid{0000-0002-1942-2831}
\affiliation{%
  \institution{University of Southern California}
  \city{Los Angeles}
  \state{CA}
  \postcode{90089}
  \country{USA}
}
\email{emiliofe@usc.edu}

\author{Luca Luceri}
\orcid{0000-0001-5267-7484}
\affiliation{%
  \institution{University of Southern California}
  \city{Los Angeles}
  \state{CA}
  \postcode{90089}
  \country{USA}
}
\email{lluceri@isi.edu}


\renewcommand{\shortauthors}{Chen and Ye et al.}

\begin{abstract}
Despite widespread concerns about the risks of AI-generated content (AIGC) to the integrity of social media discourse, little is known about its scale and scope, the actors responsible for its dissemination online, and the user responses it elicits. In this work, we measure and characterize the prevalence, spreaders, and emotional reception of AI-generated political images. Analyzing a large-scale dataset from Twitter/$\mathbb{X}$ related to the 2024 U.S. Presidential Election, we find that approximately 12\% of shared images are detected as AI-generated, and around 10\% of users are responsible for sharing 80\% of AI-generated images. AIGC superspreaders---defined as the users who not only share a high volume of AI-generated images but also receive substantial engagement through retweets---are more likely to be $\mathbb{X}$ Premium subscribers, have a right-leaning orientation, and exhibit automated behavior. 
Their profiles contain a higher proportion of AI-generated images than non-superspreaders, and some engage in extreme levels of AIGC sharing.
Moreover, superspreaders' AI image tweets elicit more positive and less toxic responses than their non-AI image tweets. This study serves as one of the first steps toward understanding the role generative AI plays in shaping online socio-political environments and offers implications for platform governance.
\end{abstract}


  

\maketitle

\section{Introduction}

Generative artificial intelligence (AI) technologies are increasingly mediating our online interactions on social media. From content moderation bots \cite{jhaver2019human} and synthetic personas \cite{ferrara2024charting} to AI-generated news and tweets \cite{kreps2022all}, AI is assisting, augmenting, or even replacing human contributions \cite{sundar2022rethinking}, creating a ``synthetic reality'' where human and AI actors coexist in our digital environment \cite{ferrara2024genai}. The advancements and proliferation of generative AI applications like ChatGPT, DALL·E, and Midjourney have amplified both the quantity and quality of AI-generated content (AIGC) online, while intensifying concerns about information credibility and authenticity \cite{lee2020authenticity}, model bias \cite{liang2021towards, ferrara2024butterfly}, social trust \cite{epstein2023art}, and potential nefarious applications \cite{minici2024uncovering}. 

Generative AI democratizes visual production by enabling the automatic creation of realistic, complex images from user imagination \cite{epstein2023art}. However, in high-stakes contexts like democratic elections, these images can project biases or be manipulated. For example, deepfakes may portray political figures saying or doing things they never did \cite{campbell2022preparing,ferrara2024charting,barari2021political}, and AI-generated memes can be weaponized to stigmatize candidates \cite{chang2024generative,minici2024uncovering}. Misuse of AI-generated images raises two primary concerns: the spread of misinformation and their impact on audience responses \cite{ferrara2024charting}. First, research consistently shows that a small fraction of individuals---also known as superspreaders---are responsible for the majority of questionable information shared on social media \cite{deverna2024identifying,nogara2022disinformation,baribi2024supersharers}. These superspreaders are more likely to be automated accounts or bots \cite{deverna2024identifying, shao2018spread}, and are often involved in the amplification of misinformation through botnets \cite{coeckelbergh2023democracy} or coordinated activities \cite{cinus2024exposing}. While AIGC clearly represents a new form of inauthentic content, it is still unknown whether superspreaders of AI-generated content display similar characteristics to misinformation superspreaders, yet, to the best of our knowledge, no prior work has examined this prolific group of users. Second, research also show that images convey information more effectively and evoke stronger emotional responses than texts \cite{li2020picture}, suggesting AI‐generated visuals may pose unique challenges. However, a systematic study of user responses to AI‐generated images on social media is still lacking.


To address this research gap, we leverage a comprehensive dataset collected from $\mathbb{X}$/Twitter during the 2024 U.S. Presidential Election period to characterize the behavior of users who share AI-generated images\footnote{AIGC and AI-generated images are used interchangeably throughout this paper.} and to measure emotional reception of AIGC over the three months leading up to the election. We initiate our analysis by exploring fundamental questions about the prevalence of AIGC and the identification of users driving its dissemination. To this end, we leverage GPT-4o to detect AI-generated images and manually validate the outputs to ensure robust and reliable classification. Through this process, we identify a set of 128 superspreaders of AI-generated images, employing established metrics of online influence \cite{deverna2024identifying}. We then characterize the behavior of these superspreaders, focusing on factors such as political affiliation, premium account status, the proportion of AIGC shared, and their bot-like behavior. Finally, we examine user responses to AI-generated images across various dimensions, including emotional tone and levels of toxicity. 

\subsection*{Contributions of this work}
Guided by our motivation to examine the \textit{prevalence}, \textit{spreaders}, and \textit{emotional reception} of AI-generated political images on $\mathbb{X}$, we formulate the following research questions (RQs):

\begin{itemize}
    \item[\textbf{RQ1:}] \textit{What is the prevalence and concentration of AI-generated images on $\mathbb{X}$?} 
    \item[\textbf{RQ2:}] \textit{What are the characteristics and sharing behaviors of AIGC superspreaders?}
    \item[\textbf{RQ3:}] \textit{How do users respond to AI-generated images on $\mathbb{X}$?}
\end{itemize}


This study serves as an initial step toward understanding how generative AI technologies shape social media dynamics in political discussions. 
As one of the first investigations using real-world data to assess the prevalence of AIGC on social media, we reveal that approximately 12\% of images within the online discourse on $\mathbb{X}$ related to the 2024 U.S. Presidential Election are AI-generated. Notably, a small fraction of users dominate AIGC dissemination, with around 10\% of image spreaders account for 80\% of the shared AI-generated images. We identify and characterize the behaviors of AIGC superspreaders, noting that they are more likely to have a right-leaning political orientation, subscribe to $\mathbb{X}$ Premium, and exhibit stronger bot-like behaviors. 
Furthermore, users tend to adopt a more positive tone and exhibit lower levels of toxicity in response to AI-generated images. We hope our findings pave the way for further research on the role of AIGC in social media, offering valuable insights for platform governance, policy-making, and public awareness of generative AI's growing impact on online sociopolitical landscapes.

\section{Related Work}

\subsection{AI-Generated Images on Social Media}
Generative AI has redefined the boundaries of visual content production. On one hand, it enables people to translate their imagination into highly customized and expressive visual content \cite{antony2025id, epstein2023art}. On the other, AI-generated images may embed biases \cite{sun2024smiling} originating from skewed training data \cite{leavy2020data} or maliciously crafted prompts. When disseminated on social media platforms, such biases can be reproduced or even amplified, potentially reinforcing harmful stereotypes, distorting public discourse, and exacerbating existing social inequalities. However, due to challenges in detecting AIGC and limited real-world datasets, only a few studies have explored its presence and impact on social media. Research has examined the prevalence and misuse of GAN-generated visuals for inauthentic activities \cite{yang2024characteristics,ricker2024ai}, the role of synthetic content like political memes in shaping discourse during elections \cite{chang2024generative,minici2024uncovering}, the use of AI-generated images to gain profit on Facebook \cite{diresta2024spammers}, and the broader influence of AIGC on platform dynamics, e.g., content creation and consumption patterns \cite{wei2024understanding}. 

Despite these contributions, existing studies lack a comprehensive characterization of AI-generated images on social media. Many are constrained to a narrow set of generative models, such as GANs \cite{yang2024characteristics,ricker2024ai}, or depend on explicit hashtags to identify labeled AIGC \cite{wei2024understanding}, thereby excluding \textit{unlabeled} content that may be misleading or harmful. Manual identification methods \cite{chang2024generative}, while useful, are neither scalable nor consistently accurate. Together, these limitations impede a full understanding of the AIGC landscape on social media platforms.

\subsection{Superspreaders of Online Information}
Superspreaders, also referred to as supersharers, are users who disproportionately contribute to the spread of specific types of content on social media, such as low-credibility or fake news. They play a unique role in shaping online information ecosystems by generating content that garners substantial reach and engagement. Studies have consistently found that a small fraction of superspreaders accounts for the majority of misinformation shared online \cite{nogara2022disinformation,deverna2024identifying}. For instance, during the 2016 U.S. Presidential Election, just 0.1\% of Twitter users were responsible for nearly 80\% of fake news shared, with similar patterns observed during the 2020 Election \cite{grinberg2019fake, baribi2024supersharers, guess2018selective}.  Superspreaders are typically identified using various metrics, such as k-core decomposition to assess centrality within the network, the sum of nearest neighbors' degrees to evaluate local influence, and the \textit{h}-index to measure the volume and reach of their shared content \cite{pei2014searching, deverna2024identifying}. Superspreaders are not limited to bots or automated accounts but frequently include politically active individuals and pundits with large followings \cite{guess2019less, baribi2024supersharers}. Recent work has also shown that $\mathbb{X}$'s recommendation algorithm amplifies superspreaders like political commentators and partisan influencers \cite{ye2024auditing}. 

While most studies have focused on superspreaders of misinformation, the rise of generative AI introduces new complexities and raises open questions. AIGC's potential to mimic diverse styles and create realistic and complex content at scale raises concerns about its integration into existing misinformation networks \cite{ferrara2024genai}. Superspreaders of AIGC may exploit these affordances and automation capabilities to amplify their influence further. This study, therefore, builds on prior work by investigating AIGC superspreaders during the 2024 U.S. Presidential Election.

\subsection{Emotional Reception of AIGC}
Understanding emotional reception of AI-generated images is both urgent and important. Visual content evokes stronger emotions than text \cite{li2020picture} and can more powerfully influence attitudes \cite{sundar2021seeing}, so it is essential to investigate how users perceive and engage with AI-generated visuals. Most prior work has relied on controlled experiments across diverse contexts, yielding mixed findings. In artwork evaluation, for instance, individuals report lower emotional engagement when they believe a piece is AI-generated \cite{agudo2022assessing}, and human-created art consistently elicits stronger emotions than AI-created art \cite{demmer2023does}. When looking into specific emotions, AI-generated images have been shown to provoke more negative reactions \cite{mehta2024emotional}. Nevertheless, in the context of architectural design, AI images effectively convey joy but poorly transmit negative emotions \cite{zhang2024decoding}.  

However, these experimental settings do not reflect the rapid, emotionally charged, and often polarized environment of social media. Few studies have explored how AI-generated images are received in real-world social feeds, where they are encountered, shared, and interpreted under naturalistic conditions. Addressing this gap is critical for understanding the public impact of AI-generated images.

\section{Methodology}

\begin{figure*}
    \centering
    \includegraphics[width=0.72\linewidth]{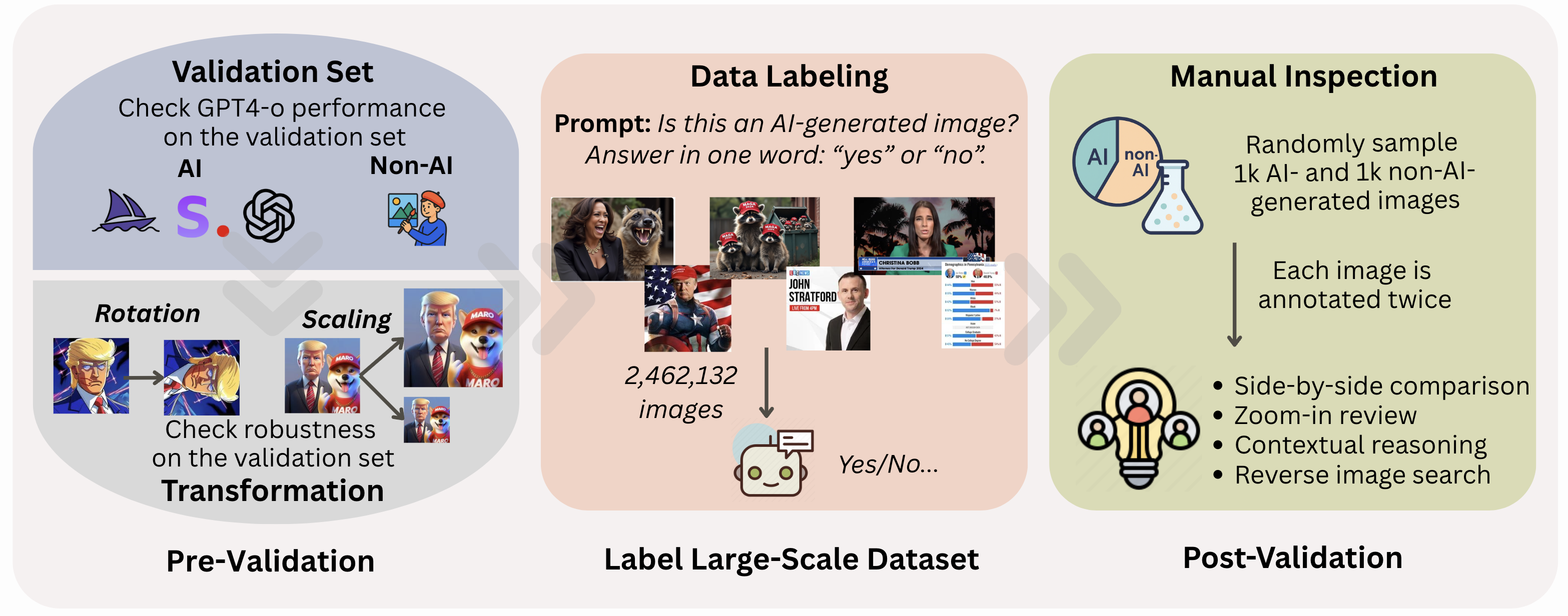} 
    \caption{Pipeline of the detection and validation of AI-generated images.} 
    \label{pipeline} 
\end{figure*}

\subsection{Data Collection and Curation}

We leverage an existing dataset of tweets related to the 2024 U.S. Election \cite{balasubramanian2024public}. The dataset is generated by querying targeted keywords related to political figures, events, and emerging issues of the Presidential Election to retrieve data effectively. In this study, we analyze data spanning a three-month period leading up to the election, from July 1, 2024, to September 30, 2024. The resulting dataset includes 2.5M images shared by 414K spreaders. Given that our analysis centers on characterizing the behavior of users sharing AI-generated content during the observation period, we first identify and quantify these AIGC spreaders. Users who share at least one AI-generated image are denoted as \textit{AI image spreaders} (details on the detection of AIGC can be found in Section \ref{sec:AIGC detection}). This yields approximately 88K AI image spreaders. 

Next, to address RQ2, we apply a filtering process to identify a subset of \textit{target users}. To ensure that each user has sufficient data points to analyze their sharing behavior and received engagement, two criteria are applied. The first criterion sets a minimum threshold for the total number of tweets a user has posted during the observed time period. AI image spreaders who have posted fewer than five tweets with at least one image in each are excluded from the analysis. Second, an AI image spreader must have shared at least one retweeted post containing AIGC, i.e., \textit{h}-index $>$ 0 (details on the \textit{h}-index are provided in Section \ref{sec:superspreaders}). These criteria collectively ensure a robust and meaningful dataset for evaluating user activity and influence. Finally, we retrieve a subset of target users consisting of 12,898 AI image spreaders. Subsequent analyses are based on this pool of target users.

\subsection{AI-Generated Image Detection}
\label{sec:AIGC detection}

With the proliferation of AI-generated content on platforms like $\mathbb{X}$, accurately detecting such images has become increasingly challenging. Traditional deep learning models often struggle to distinguish AI-generated images \cite{borji2022good}. In contrast, recent work demonstrates that large language models (LLMs) such as GPT-4o, when guided by carefully crafted prompts, offer robust multimodal detection capabilities \cite{chen2023gpt,chang2024generative,brin2024assessing}. Drawing on these advances, we employ GPT-4o to identify AI-generated images in this study. Figure~\ref{pipeline} presents an overview of our detection and validation pipeline.

We leverage GPT-4o to classify images as AI-generated or non-AI-generated. To evaluate its performance, following \citet{epstein2023online}, we construct a validation set of 2,400 images, evenly split between 1,200 AI-generated and 1,200 non-AI-generated samples. The non-AI images are randomly sampled from the LAION-400M dataset \cite{schuhmann2021laion}, while the AI-generated set comprises 400 DiffusionDB samples \cite{wang2023diffusiondb}, 400 DALL·E images collected from the subreddit \texttt{r/dalle}\footnote{\url{https://www.reddit.com/r/dalle/}}, and 400 Midjourney images crawled from the Explore page\footnote{\url{https://www.midjourney.com/explore/}}.


We extract image URLs from tweet metadata and send the following prompt to GPT-4o (temperature set to 0 to minimize variability):

\begin{quote}
\textit{Is this an AI-generated image? Answer in one word: ``yes'' or ``no''.}
\end{quote}

Of the 2,400 images to be validated, the model returns responses for 2,367 and fails on 33 (often due to sensitive‐content filters), achieving an F1‐score of 0.96. To assess robustness, we employ two additional transformations to the same 2,400 images:

\begin{itemize}
  \item \textbf{Rotation:} We split the images into three equal groups, rotating them by 90°, 180°, or 270°. GPT-4o processes 2,380 images (20 failures) and reaches an F1‐score of 0.92.
  \item \textbf{Scaling:} We resize half of the images to 200\% and the other half to 50\% of their original sizes. GPT-4o processes 2,366 images (34 failures) and achieves an F1‐score of 0.94.
\end{itemize}

These results demonstrate GPT-4o’s high accuracy and resilience to common image transformations. We then apply GPT-4o detection to our 2024 U.S. Election dataset using the OpenAI BatchAPI\footnote{OpenAI BatchAPI: \url{https://platform.openai.com/docs/guides/batch}} to optimize time and cost. We categorize each response as: (i) valid (“yes” or “no”), (ii) download failure due to content moderation, or (iii) inability to respond (e.g., “Sorry, I cannot answer this.”). Categories (ii) and (iii) are treated as invalid answers. Of the 2,462,132 images processed, approximately 90.5\% receive valid responses. Examples of detected AI-generated images are shown in \textit{Appendix}. All subsequent analyses focus on these valid detections and their associated users. \textit{The images and detection results have been uploaded to an anonymized GitHub repository, which will be made publicly available upon acceptance. } 

\subsection{Validation of AIGC Detection}
Although GPT-4o performs well on our constructed validation set, its accuracy on real-world social media images remains uncertain due to the variability and complexity of online content. To assess robustness in the wild, we adopt an annotation workflow inspired by Ricker et al. \cite{ricker2024ai} that introduces annotation guidelines and heuristic-based training for annotators, followed by validation of samples from model-labeled datasets. Following a similar approach, we randomly sample 1,000 images labeled as AI-generated and 1,000 labeled as non-AI-generated by GPT-4o from our large-scale dataset. We then provide annotators with a detailed guideline to validate model's performance in the wild.

To facilitate consistent annotation, we provide annotators with a set of heuristic strategies: \textit{(i) side-by-side comparison}, where annotators compare the current image to ground-truth-labeled examples, focusing on their visual patterns and framing; \textit{(ii) zoom-in review of visual cues}, which encourages annotators to identify common generation-related artifacts such as unnatural textures and object distortions; and \textit{(iii) contextual reasoning}, which prompts annotators to apply basic logics to check the plausibility of the scene. These techniques, which have proven effective in previous work ~\cite{ricker2024ai}, are aimed at improving both accuracy and inter-coder reliability.


Beyond these heuristics, we integrate an additional technique inspired by ~\citet{qian2023fighting}: \textit{Google reverse image search}. Their study has shown that introducing reverse image search into a digital media literacy intervention significantly increases users’ intention and ability to verify images. The underlying rationale is that this tool allows users to verify the origins and context of an image by searching with the image itself rather than a keyword. When an exact or near-duplicate match is found online, the accompanying metadata, such as early publication dates or credited painters or photographers, can serve as strong evidence that the image is non-AI-generated. By incorporating this step, we not only provide annotators with external validation cues but also align our workflow with proven strategies to enhance media discernment and validation robustness.


Following the annotation guidelines, each image is independently labeled by two different annotators. We first assess inter-coder reliability, obtaining a Cohen's $\kappa$ of 0.87, which indicates a high level of agreement. We then evaluate the model's performance by calculating its misclassification rates, yielding a false positive rate (FPR) of 6.5\% and a false negative rate (FNR) of 1.6\%. These results confirm the robustness of our detector when applied to the large-scale dataset in real-world scenarios. While the error rates are not zero, they are sufficiently low to support the validity of the subsequent analysis.

\subsection{Identifying Superspreaders of AIGC}
\label{sec:superspreaders}
Superspreaders are typically identified using influence metrics, including k-core centrality, the sum of nearest neighbors' degrees, and the engagement driven by their original content
\cite{pei2014searching,deverna2024identifying}. 
Here, we adopt the \textit{h}-index, originally designed to measure scholarly impact \cite{hirsch2005index}, and later adapted for identifying superspreaders of low-credibility content on $\mathbb{X}$ \cite{deverna2024identifying}. In our scenario, the $h$-index for a user $i$ is defined as the maximum value of $h$ such that user $i$ posted at least $h$ tweets containing AIGC, each retweeted at least $h$ times. This metric captures both the volume of AI-generated content shared and the engagement received by each post. A superspreader must meet both criteria: a high number of tweets containing AI-generated content and a significant volume of received retweets. This ensures that superspreaders are identified based on both their activity and the broader impact of their AI-generated content on the platform.

\subsection{Characterizing AIGC Superspreaders}

We characterize superspreaders of AI-generated content based on two key metrics: The \textit{AI score} quantifies the proportion of AI-generated content in a user's posts. The \textit{bot score} evaluates the probability of an account being automated based on its historical tweet activity.

\subsubsection{AI Score.}
To quantify the proportion of AI-generated content in a user's tweets, we introduce a new metric called the \textit{AI score}, defined as:
\[
r_i = \frac{A_i}{T_i},
\]
where $A_i$ is the number of tweets containing at least one AI-generated image, and $T_i$ is the total number of tweets containing images shared by the user $i$ during the observation period. This metric captures the dominance of AIGC in a user's shared posts, with higher AI scores indicating a greater reliance on AI-generated images.

\subsubsection{Bot Score.}
We use the bot score metric to assess the likelihood of an account being automated. Bot scores are calculated using the widely adopted tool Botometer \cite{ferrara2016rise}, which evaluates various account features, such as tweet content, network interactions, and posting behavior. A machine learning model then assigns a score ranging from 0 to 1, where higher scores indicate a greater likelihood of automation. Notably, Botometer relies on the Twitter V1 API \cite{yang2020scalable} to retrieve historical account data. Since this API was discontinued in 2023, we can only retrieve bot scores for accounts created prior to May 2023.

\subsection{Examining Emotional Reception of AIGC}
We aim to compare differences in user responses of AI image tweets versus non-AI image tweets shared by AIGC superspreaders. Our analysis focuses specifically on the comment section, as comments serve as an important indicator for understanding audience reception in social media research. We begin by sampling tweets from AIGC superspreaders and collecting the associated comments. 
We then assess emotional reception across two dimensions: emotion and toxicity. \textit{Emotion recognition} assigns a probability score to each comment across eleven emotion categories while the \textit{language toxicity} metric evaluates the degree of toxicity present in the comments.

\subsubsection{Curation of a Comments Dataset}
We select comments from tweets shared by AIGC superspreaders for analysis. Here, we define \textit{AI image tweets} as those in which all images are AI-generated, and \textit{non-AI image tweets} as those in which all images are non-AI-generated. This definition allows us to more clearly isolate the effects of AI-generated content and ensure the validity of the comparison. Our goal is to compare the emotional reception of AI and non-AI image tweets posted by the same AIGC superspreaders. This design serves two purposes. First, by focusing on tweets from influential accounts, we ensure that the content is likely to attract sufficient attention and generate a substantial volume of comments for analysis. Second, using the same set of superspreaders helps control for user identity, which may influence user responses independent of tweet content. This allows us to better isolate the effect of image type on audience reception.


Next, we randomly sample 300 tweets shared by the 128 AIGC superspreaders for each group (AI image tweets vs. non-AI image tweets). To ensure the sample reflects users’ sharing contribution, we first compute a sampling weight for each user based on their share of tweets within the group. Specifically, the weight for user~$i$ is defined as:
\[
w_i = \frac{n_i}{N},
\]
where $n_i$ denotes the number of tweets posted by user~$i$, and $N$ is the total number of tweets in that group. Each user is then assigned a target sample size by multiplying their weight~$w_i$ by 300 and rounding up, ensuring that every user contributes at least one tweet. For each user, we rank their tweets by the number of replies in descending order and select the top entries to prioritize tweets with higher engagement and reduce the likelihood of including tweets with zero replies. We then collect user comments associated with these sampled tweets, resulting in 16,832 comments for the AI image tweets and 27,848 comments for the non-AI image tweets.



\subsubsection{Emotion Recognition}
To conduct emotion recognition on users' comments, we utilize TweetNLP \cite{camacho2022tweetnlp}, a state-of-the-art NLP toolkit specifically designed for social media text, which is built upon transformer-based language models. We employ its emotion recognition module, which for each input comment, returns a probability distribution over eleven emotion categories, including anger, anticipation, disgust, fear, joy, love, optimism, pessimism, sadness, surprise, and trust, where each score ranges from 0 to 1 and reflects the predicted intensity of that emotion. We compute each tweet's response emotion scores by averaging the emotion scores of all its comments.

\subsubsection{Language Toxicity.}
To examine the use of toxic language among comments from AI image tweets and non-AI image tweets, we draw from the Google Jigsaw Perspective API \cite{lees2022new}, a machine learning tool designed to identify and manage harmful or abusive content on online platforms. The API assigns a toxicity score to each input comment, ranging from 0 to 1, with higher scores indicating a greater likelihood of rude or harmful language. This analysis focuses exclusively on English-language comments. We compute each tweet's response toxicity scores by averaging the toxicity scores of all its comments.

\begin{figure}[t]
    \centering
    \includegraphics[width=0.75\linewidth]{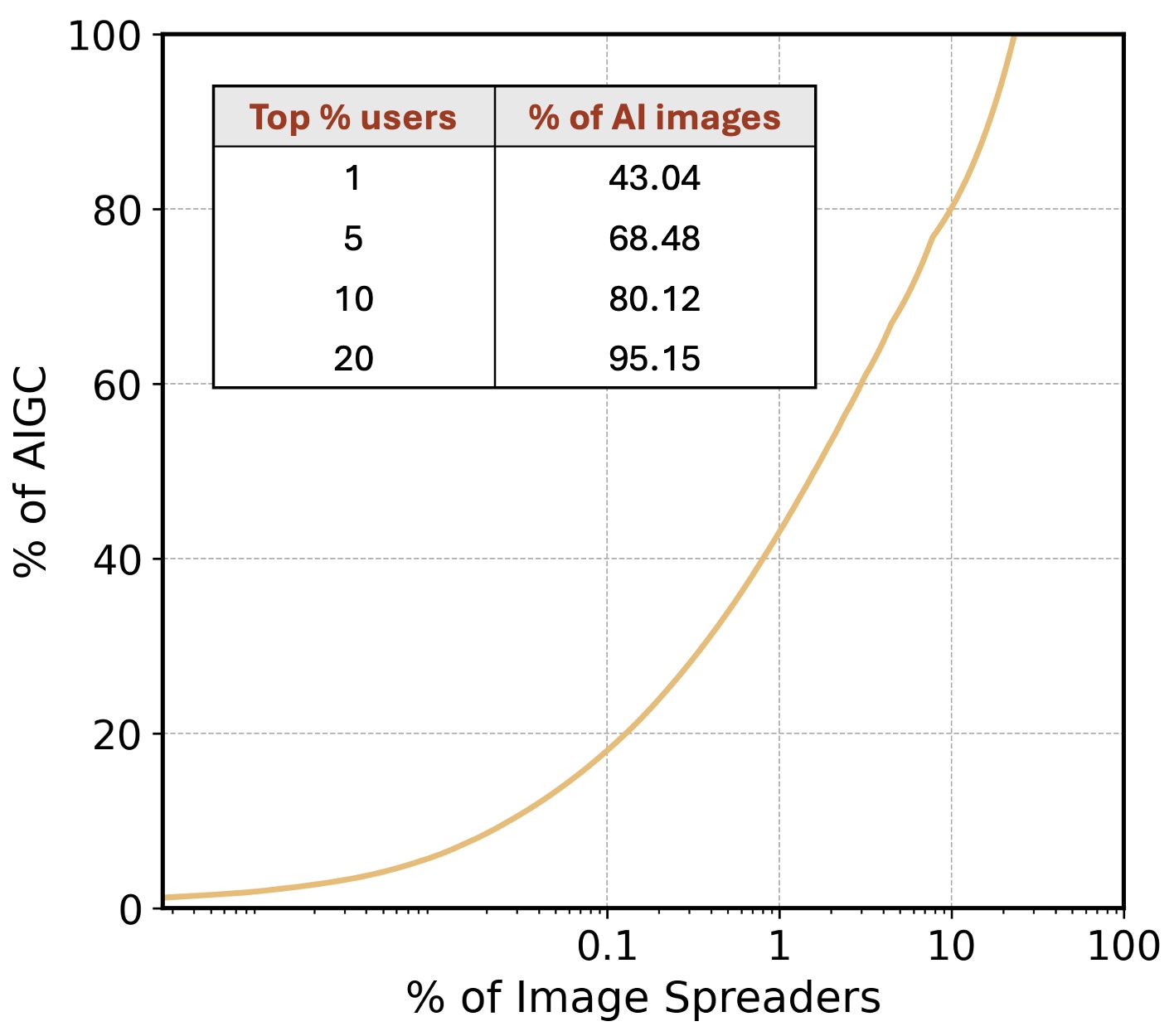} 
    \caption{Cumulative distribution function (CDF) of the total number of AI-generated images shared by image spreaders.} 
    \label{inequality}
\end{figure}

\section{Results}

\subsection{RQ1: Prevalence and Concentration of AIGC}
Investigating the prevalence of AI-generated images involves two key questions. First, what proportion of all images on $\mathbb{X}$ are AI-generated? Second, what proportion of users sharing images have shared AI-generated content?


Overall, approximately 12.33\% of images related to the 2024 U.S. Election on $\mathbb{X}$ are AI-generated according to GPT-4o classification. In terms of users sharing AIGC, among 379,025 users who share at least one image, 23.23\% are AI image spreaders. Regarding sharing inequality, Figure \ref{inequality} shows that approximately 10\% of image spreaders account for 80\% of shared AI-generated images. This suggests that the sharing of AI-generated images is highly concentrated, with a small number of users responsible for sharing a large volume of AI-generated images.

\subsection{RQ2: AIGC Superspreader Characterization}

\begin{figure*}
    \centering
    \includegraphics[width=0.73\linewidth]{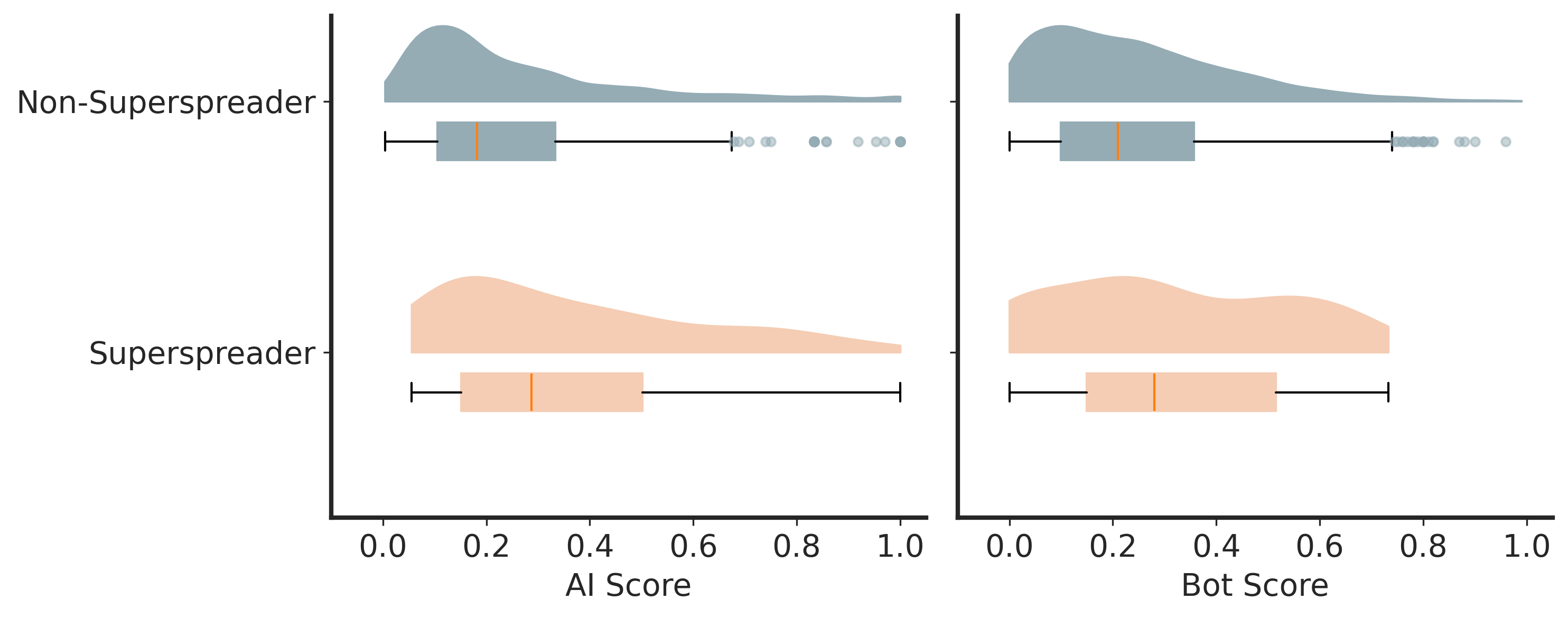} 
    \caption{AI scores (left) and bot scores (right) across user groups. According to the Mann-Whitney U test, Superspreaders exhibit significantly higher AI scores than Non-Superspreaders ($p$<.001), and significantly higher bot scores ($p$<.01).} 
    \label{ai_bot} 
\end{figure*}

In this section, we identify and characterize superspreaders within our target user group, which consists of 12,898 AI image spreaders. Using the \textit{h}-index, we refer to the top 1\% of the accounts with the highest \textit{h}-index as \textit{superspreaders}, following \citet{deverna2024identifying}. This yields 128 AIGC superspreaders and 12,770 AIGC non-superspreaders.

We first look into the political leanings of these superspreaders and their account verification statuses. To further explore the unique characteristics of AIGC superspreaders compared to non-superspreaders, we examine their sharing behaviors using two key metrics: AI score and bot score. Group-wise comparisons are conducted using the Mann-Whitney U test, with significant differences reported using \textit{p}-values where applicable.

\subsubsection{Political Affiliation.} 
One way to characterize these AIGC superspreaders is by examining their political affiliation. To achieve this, two authors manually inspect the 128 accounts.
The annotation process is conducted in two steps. First, the annotators assess whether the user’s profile description indicates a clear political affiliation. Second, if an affiliation is present, they determine whether the user supports a specific political party. After the initial round of annotations, the two annotators achieve a Cohen’s $\kappa$ of 0.91, indicating a high level of inter-coder reliability. For the remaining disagreements, a third annotator is involved, and the disagreements are resolved through a majority vote among the three annotators. To further validate our annotation of superspreaders’ partisanship, we collect all tweets they post in our dataset and calculate the frequency of hashtags. Table \ref{tab:top_hashtags} presents the top five hashtags most frequently used by left- and right-leaning users\footnote{In the U.S. political context, left-leaning refers to Democratic-leaning and right-leaning to Republican-leaning.}.

\begin{table}[t]
\centering
\caption{Top 5 hashtags (and their occurrences) used by left- and right-leaning superspreaders.}
\begin{tabular}{ll}
\toprule
\textbf{Left-leaning} & \textbf{Right-leaning} \\
\midrule
\#harriswalz2024 (123)       & \#maga (1,290) \\
\#demsunited (85)        & \#trump2024 (1,138) \\
\#proudblue (54)        & \#fjb (285) \\
\#usdemocracy (53)  & \#ifbap (143) \\
\#demvoice1 (46) & \#demvoice1 (113) \\
\bottomrule
\end{tabular}
\label{tab:top_hashtags}
\end{table}

As shown in Table~\ref{tab:top_hashtags}, users in the left-leaning group predominantly use hashtags that emphasize party identity and collective action, such as \#demsunited, \#demvoice1, and \#proudblue, as well as support for specific candidates (e.g., \#harriswalz2024). In contrast, right-leaning superspreaders frequently employ hashtags associated with Donald Trump and his political messaging, including \#maga and \#trump2024, which appear with notably high frequency. Additionally, right-leaning users express anti-Biden sentiment using hashtags such as \#fjb. Notably, a few hashtags like \#demvoice1 also appear in the right-leaning group. Based on a manual inspection of randomly sampled tweets, these hashtags are often used in sarcastic contexts. However, the frequency is substantially lower than top hashtags like \#maga and \#trump2024, suggesting that cross-group hashtag usage is marginal and does not undermine the overall partisan distinctiveness.

The results show that among the 128 AIGC superspreaders, 88 have clear political affiliations, with 64 identified as right-leaning and 24 as
left-leaning, while 40 users show no clear affiliation. We conclude that among users with clearly established partisanship, a substantial majority (72.2\%) is right-leaning. Moreover, around 31.3\% of users lack clear political affiliations.

\subsubsection{Premium Account Status.} 
Another perspective to describe superspreaders is to examine the proportion of \textit{$\mathbb{X}$ Premium subscribers}\footnote{$\mathbb{X}$ Premium:\url{https://help.x.com/en/managing-your-account/about-x-verified-accounts}}. $\mathbb{X}$ Premium subscribers (users with a blue check mark) gain additional features, such as tweet editing, longer posts, enhanced analytics, and priority in replies and searches. Results show 82.8\% of AIGC superspreaders are \textit{$\mathbb{X}$ Premium subscribers}, compared to 48.7\% among non-superspreaders, suggesting a potential association between premium subscription status and high-volume dissemination of AIGC.

\subsubsection{AI Score.} 
The AI score represents proportions of AI-generated content shared by a user. Figure \ref{ai_bot} (left) shows the AI scores for AIGC superspreaders and non-superspreaders. Results show that AIGC superspreaders have significantly higher AI scores than non-superspreaders ($p < .001$),
indicating greater involvement in sharing AI-generated content. This outcome is expected but not guaranteed, as hyperactive accounts do not always garner high engagement \cite{luceri2021down}. 
Moreover, the fat-tailed distribution among superspreaders highlights the prevalence of a relevant fraction of users with high AI scores exhibiting extreme AIGC-sharing behaviors. 

\subsubsection{Bot Score.} 
Although the analysis using Botometer is limited to accounts created prior to the discontinuation of the Twitter API, a significant portion of our user base was successfully evaluated, indicating that the majority of the accounts analyzed are not newly created. In particular, we assess 99 (77.3\%) AIGC superspreaders and 10,415 (81.6\%) AIGC non-superspreaders.

Figure \ref{ai_bot} (right) presents the bot scores for AIGC superspreaders and non-superspreaders. First, we can observe that superspreaders are significantly more likely to exhibit automated behavior than non-superspreaders ($p < .01$). Second, the distribution of superspreaders is distinctly bimodal, with prominent peaks around 0.2 and 0.6, suggesting a conspicuous number of automated accounts employed in the diffusion of AIGC. Note that a high bot score (e.g., larger than 0.5) is indicative of bot-like behavior \cite{luceri2019evolution}.

\begin{figure*}[t]
    \centering
    \includegraphics[width=0.84\textwidth]{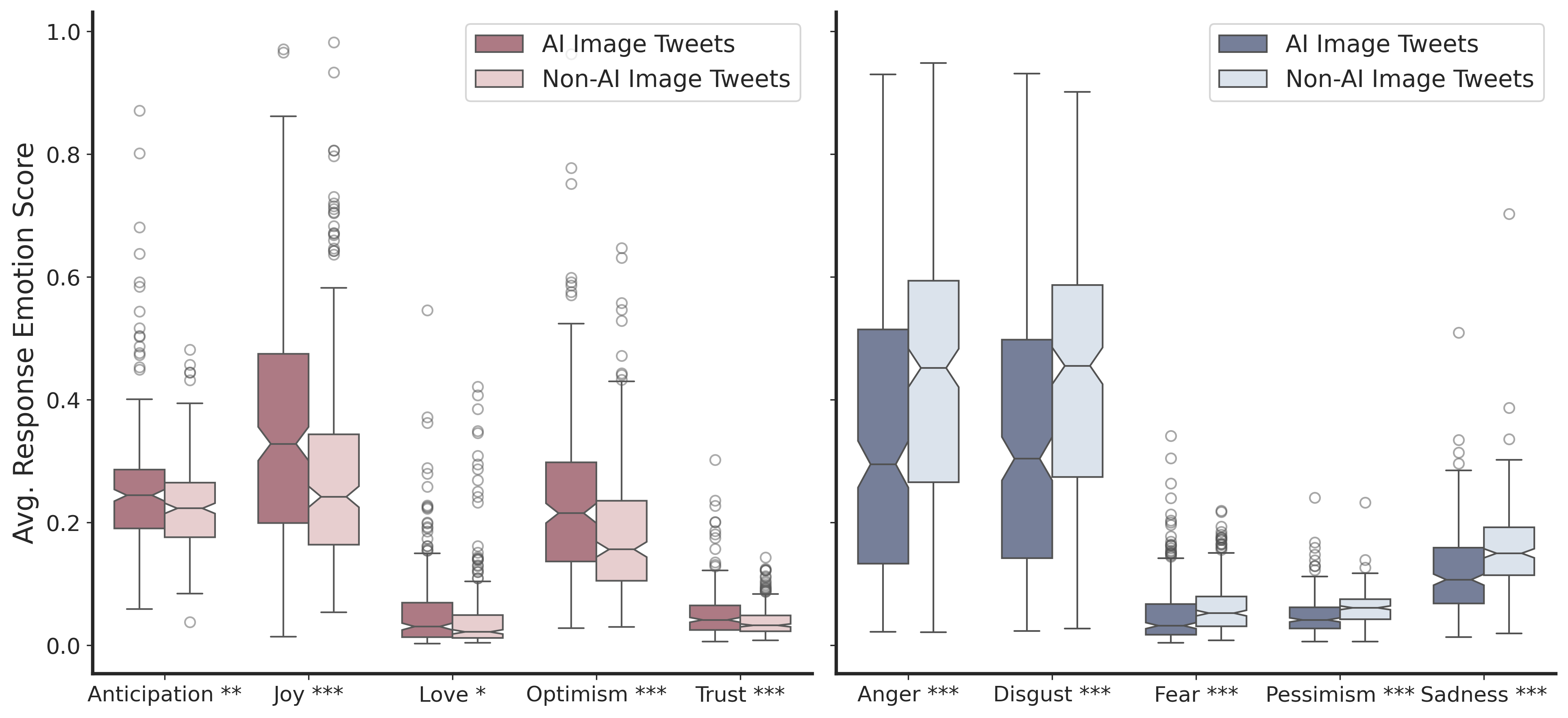}
    \caption{Comparison of emotion scores across AI image tweets vs. non-AI image tweets, both shared by AIGC superspreaders. We apply the Mann-Whitney U test to compare the two tweet groups across emotion categories. All categories shown are statistically significant (*\,$p$\,$<$\,.05, **\,$p$\,$<$\,.01, ***\,$p$\,$<$\,.001); The emotion \textit{surprise} is not shown as it is not significant.}
    \label{emotion} 
\end{figure*}

\subsubsection{Summary.} Among the AIGC superspreaders, a higher proportion of users align with conservative views. Premium accounts are more prevalent among superspreaders. When characterizing the behaviors of AIGC superspreaders, three key findings emerge: (i) AIGC superspreaders tend to share a higher proportion of AI-generated content in their profiles compared to non-superspreaders; (ii) A substantial number of AIGC superspreaders exhibit a strong reliance on AI-generated images in their sharing activity, reflecting extreme content dissemination behaviors; (iii) AIGC superspreaders demonstrate a higher degree of bot-like behavior than non-superspreaders.

\subsection{RQ3: Emotional Reception of AIGC}
In this section, we compare emotional reception of AI image tweets and non-AI image tweets posted by AIGC superspreaders. The analysis focuses on two dimensions: First, we examine \textit{response emotion} across the two groups using TweetNLP's emotion classification module; second, we assess \textit{response toxicity} between the two groups of tweets. Group-wise comparisons are conducted using the Mann–Whitney U test, and statistically significant differences are reported with corresponding \textit{p}-values where applicable.

\subsubsection{Response Emotion}
We assess the probability distribution across 11 emotion categories for each comment and compute the average emotion scores for each tweet. Figure~\ref{emotion} reveals consistent and statistically significant differences in emotional responses between AI image tweets and non-AI image tweets across nearly all emotion categories, except for \textit{surprise}. An interesting observation is that AI image tweets achieve higher scores than non-AI image tweets across all positive emotions (\textit{anticipation}, \textit{joy}, \textit{love}, \textit{optimism} and \textit{trust}). In contrast, non-AI image tweets score higher than AI image tweets across all negative emotions (\textit{anger}, \textit{disgust}, \textit{fear}, \textit{pessimism} and \textit{sadness}). The consistently greater arousal of positive emotions in response to AI image tweets points to a unique emotional effect that this type of content exerts on users.


\subsubsection{Response Toxicity}
\begin{figure}[t]
    \centering
    \includegraphics[width=0.77\linewidth]{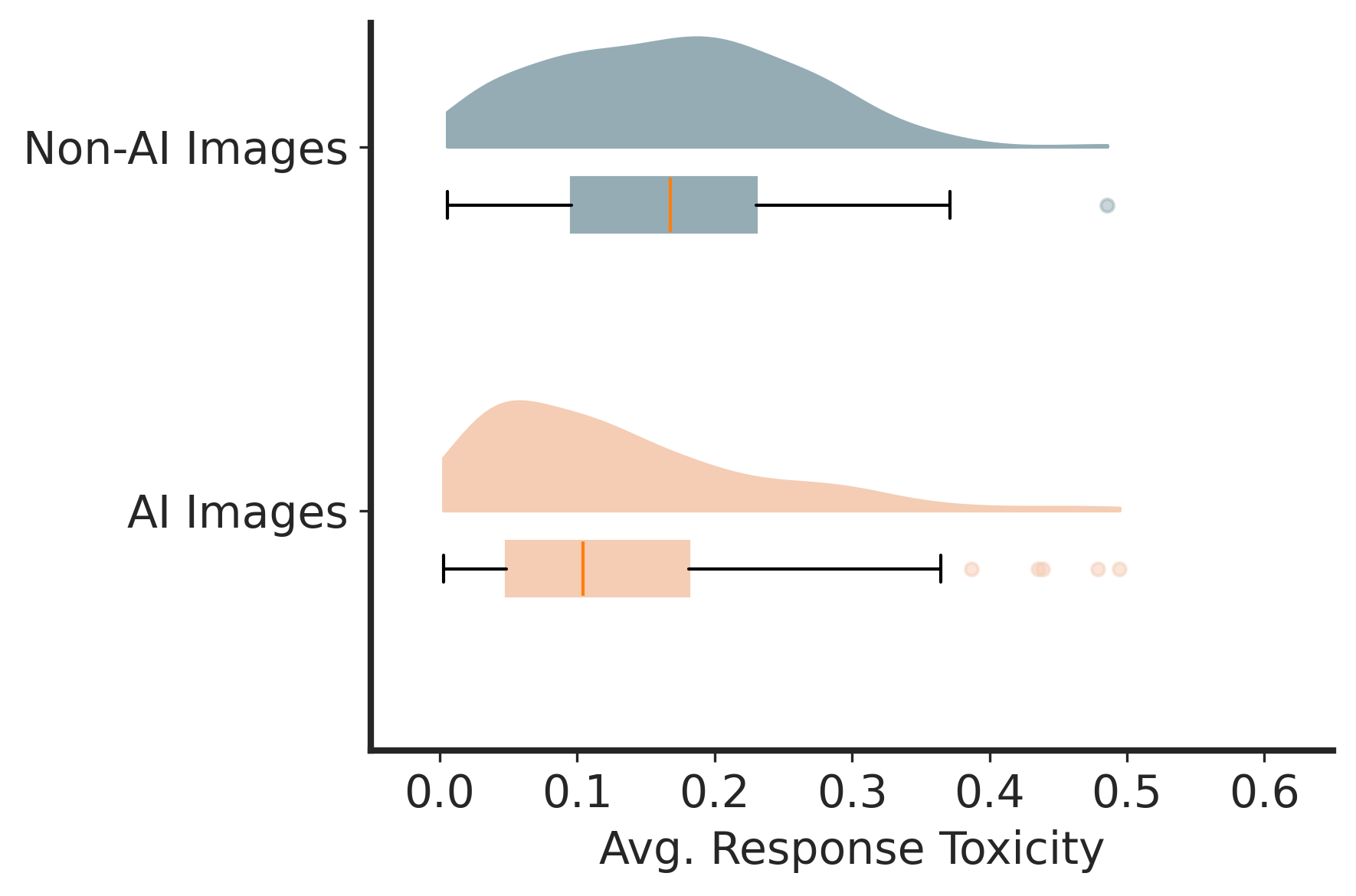} 
    \caption{Comparison of toxicity scores between AI and non-AI image tweets shared by AIGC superspreaders. The Mann-Whitney U test indicates a significant difference ($p$<.001)}
    \label{toxicity}
    \vspace{-0.2cm}
\end{figure}

For the level of toxicity, the Mann-Whitney U test reveals a significant difference in toxicity scores between responses to AI image tweets and non-AI image tweets. As shown in Figure~\ref{toxicity}, comments on non-AI image tweets exhibit significantly higher levels of toxicity ($p < .001$), suggesting a greater prevalence of hostile or harmful language. This finding is also consistent with the previous discovery indicating that non-AI image tweets are more likely to elicit negative emotions.

\subsubsection{Summary}
Our analysis reveals clear distinctions in the emotional reception of AI versus non-AI image tweets. AI image tweets tend to evoke significantly more positive emotions while non-AI image tweets are more likely to elicit negative ones. This suggests a potential affective advantage of AI-generated content in fostering positive engagement. Moreover, comments on non-AI image tweets exhibit significantly higher toxicity scores, indicating a greater prevalence of hostile or harmful language compared to responses to AI image tweets. 

\section{Discussion \& Conclusions}
This study, grounded in the analysis of approximately 2.5 million images, investigates the prevalence and concentration of AI-generated content (AIGC), provides an in-depth characterization of AIGC superspreaders, and examines the emotional reception of AIGC. Four key insights emerge from our research.

\paragraph{Finding 1: AIGC sharing is prevalent and unequal.}
Our analysis reveals that approximately 12\% of images are deemed AI-generated and that 10\% of image spreaders account for 80\% of all AI-generated images. Generative AI applications have significantly lowered the barriers to creating and sharing customized content, such as political memes \cite{chang2024generative}. While this trend highlights the growing presence of AIGC on social media, it also raises concerns about misuse, including attacks on political figures \cite{chang2024generative}, fabrication of events, and dissemination of misinformation \cite{ferrara2024charting}, underscoring the urgent need for robust detection mechanisms and control strategies. Besides, the skewed distribution of sharing activity highlights the outsized role of a small group of users in AIGC dissemination. Similar concentration patterns have been observed in misinformation sharing, where a small fraction of users disproportionately spread low-credibility content \cite{grinberg2019fake,baribi2024supersharers,deverna2024identifying,nogara2022disinformation}. For instance, \citet{grinberg2019fake} shows that 0.1\% of users share roughly 80\% of fake news. However, AIGC sharing appears less skewed, suggesting a broader base of users is involved in its dissemination and potentially greater exposure to AI-generated content. These findings call for further research into the dynamics of AIGC diffusion to gain deeper insights into its reach, engagement, and influence relative to misinformation.

\paragraph{Finding 2: AIGC superspreaders are more likely to be bots.}
Our results show that AIGC superspreaders have significantly higher bot scores compared to non-superspreaders, indicating that they are more likely to be automated accounts. This aligns with prior studies showing that bots play a crucial role in diffusing low-credibility content \cite{shao2018spread}, highlighting similarities between the dissemination patterns of AIGC and misinformation. However, an intriguing discrepancy also emerges. Unlike \citet{shao2018spread}, which reports a single peak in the bot score distribution around 0.2, we discover that the distribution of bot scores among AIGC superspreaders seems to exhibit a binomial distribution, with two peaks at around 0.2 and 0.6. This might surface the potential coordinated activity of a subset of bots specifically targeting the dissemination of visual AIGC. Therefore, future research could look into the strategies, motivations, and coordinated actions underlying bot-driven AIGC diffusion.

\paragraph{Finding 3: AIGC sharing can be heterogeneous and extreme.}
Our analysis reveals a broad range of AI scores among users sharing AI-generated images, suggesting great variability in how users adopt and share this content. The diversity of AI scores reflects a wider spectrum of behaviors among AI image spreaders, ranging from sporadic use to highly frequent generation of AI-generated visuals. Notably, the distribution of AI scores among AIGC superspreaders displays a ``fat tail,'' indicating the presence of a small yet substantial fraction of users who engage in extreme sharing behaviors. These outliers play a disproportionate role in sharing and driving engagement with their AI-generated images on social media, compared to the majority of users who share AIGC at a more moderate level.

\paragraph{Finding 4: AIGC elicits more positive and less toxic responses.}
Our findings demonstrate that comments on superspreaders' AI image tweets are more positive and less toxic than non-AI image tweets. This pattern aligns with previous research showing that AI-generated images can effectively convey positive emotions, such as joy \cite{zhang2024decoding}. One possible explanation is that AI-generated content may introduce an entertainment framing of political messages, which has been shown to reduce hostility and aggression in public discourse by creating a more playful and relaxed atmosphere \cite{lee2022power}. This interpretation is further supported by a close manual inspection of sampled AI-generated images (see in \textit{Appendix}), which often exhibit playful, humorous, or creatively exaggerated elements that help diffuse tension in otherwise contentious discussions. Also, future research is encouraged to further examine the content of images that elicit positive emotions, with a focus on identifying common visual attributes and thematic elements that are particularly effective in triggering specific emotions.

\paragraph{Limitations and Future Work}
We acknowledge several limitations in our research, which will guide future research. First, while this study focuses on analyzing individual superspreaders, future research could investigate the collective behaviors of coordinated superspreaders, offering a broader understanding of their networked activity and impact on vulnerable population. Second, while this study centers on characterizing AIGC superspreaders and analyzing user responses, future research could extend this line of inquiry by examining the content-level features of AIGC. Identifying the shared attributes of AI-generated content that are more effective in driving user engagement and eliciting emotional responses would offer valuable insights into the persuasive power and communicative potential of AIGC. Lastly, this study is situated within the political context of $\mathbb{X}$; future research could extend and compare these findings to other platforms and different high-stakes scenarios, such as public health crises or misinformation campaigns.


\paragraph{Ethical Statement}
This study investigates the behaviors of 
AIGC superspreaders and user responses to AIGC on social media, aiming to inform platform governance and public awareness. To mitigate potential societal harms, we anonymize all user data and avoid releasing granular user-level findings that could be exploited to enhance malicious content dissemination. We acknowledge the potential misuse of our findings, such as stigmatizing specific users or enabling more effective content manipulation, and have taken steps to ensure responsible data handling, access control, and reproducibility while adhering to ethical research standards.


\balance
\bibliographystyle{ACM-Reference-Format}
\bibliography{reference}

\newpage
\onecolumn
\section*{Appendix}
\subsection*{A. Examples of Detected AI-Generated Images}

\begin{figure}[h]
    \centering
    \includegraphics[width=0.4\linewidth]{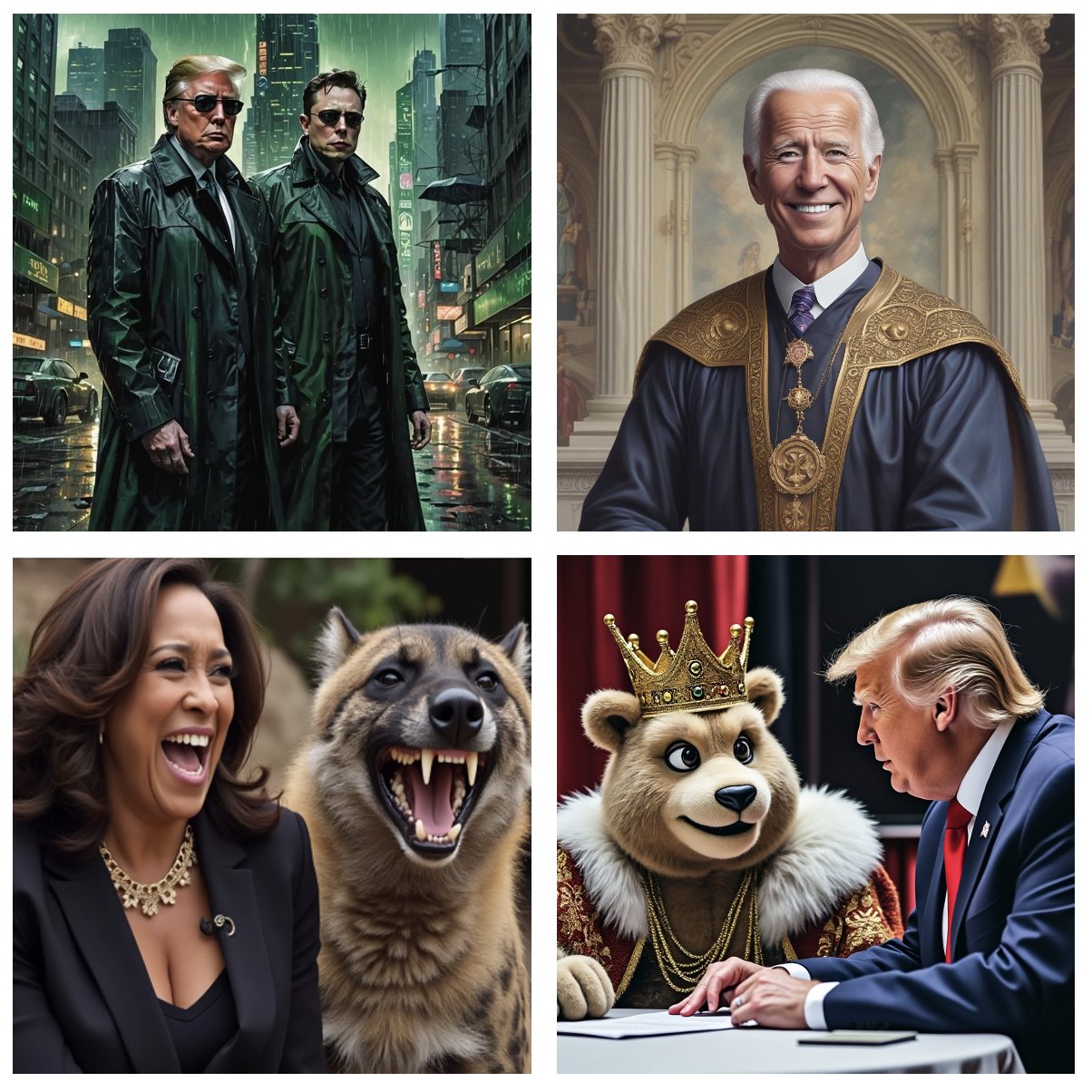} 
    \includegraphics[width=0.4\linewidth]{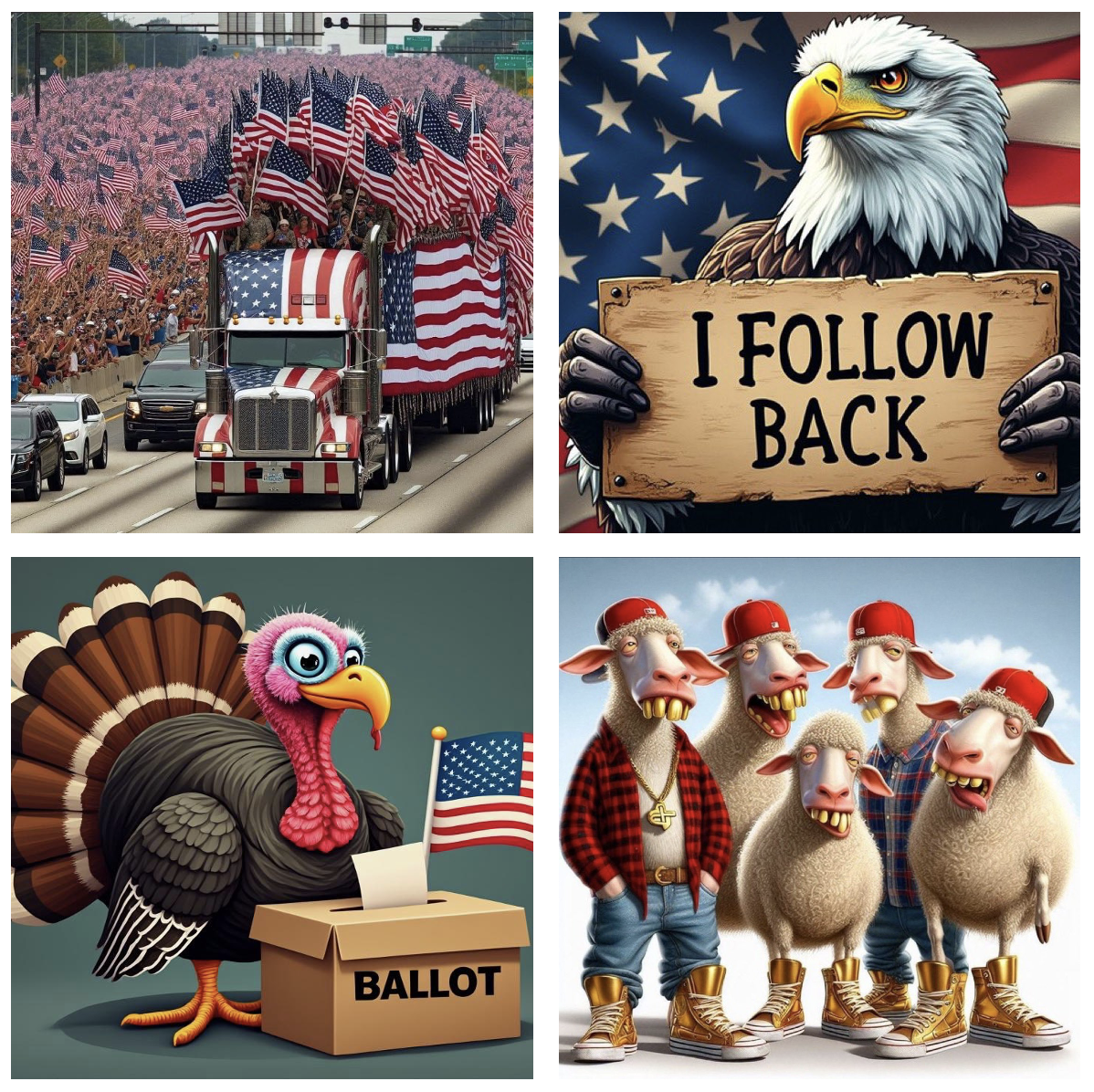} 
    \caption{Examples of images detected as AI-generated in our dataset.}
    \label{example_1}
\end{figure}

Examples shown in Figure~\ref{example_1}, illustrate typical AI-generated political images identified by our detection pipeline. By presenting representative cases, we aim to give the reader an intuitive sense of the visual cues our annotators and automated model rely upon when classifying content as AI-generated.

\end{document}